\documentclass[aps,prb,reprint,superscriptaddress,floatfix]{revtex4-2}
\usepackage{amssymb}
\usepackage{physics}
\usepackage{graphicx}
\usepackage{amsmath}
\usepackage{amsthm}
\usepackage[detect-none]{siunitx}
\usepackage{amsfonts}
\usepackage[T1]{fontenc}
\usepackage{array}
\usepackage{multirow}
\usepackage{color}
\usepackage{esint}
\usepackage{bm}
\usepackage{color}
\usepackage{bbm}
\usepackage{hyperref}
\usepackage{babel}

\newcommand{\moire}{moir\'{e}}
\usepackage{hyperref}
\usepackage[normalem]{ulem}
\hypersetup
{	colorlinks,%
	citecolor=green,%
	linkcolor=red,%
	urlcolor=blue%
}
\allowdisplaybreaks[4]

\begin{document}

\global\long\def\id{\mathbbm{1}}
\global\long\def\ui{\mathbbm{i}}
\global\long\def\ud{\mathrm{d}}

\title{Fractional quantum anomalous Hall effect in rhombohedral multilayer graphene with a strong displacement field}

\author{Ke Huang}
\affiliation{Department of Physics, City University of Hong Kong, Kowloon, Hong Kong SAR}
\author{Sankar Das Sarma}
\affiliation{Condensed Matter Theory Center and Joint Quantum Institute, Department of Physics, University of Maryland, College Park, Maryland 20742, USA}
\author{Xiao Li}
\email{xiao.li@cityu.edu.hk}
\affiliation{Department of Physics, City University of Hong Kong, Kowloon, Hong Kong SAR}

\date{\today}

\begin{abstract}
We investigate the fractional quantum anomalous Hall (FQAH) effect in rhombohedral multilayer graphene (RnG) in the presence of a strong applied displacement field. 
We first introduce the interacting model of RnG, which includes the noninteracting continuum model and the many-body Coulomb interaction. 
We then discuss the integer quantum anomalous Hall (IQAH) effect in RnG and the role of the Hartree-Fock approach in understanding its appearance. 
Next, we explore the FQAH effect in RnG for $n=\numrange{3}{6}$ using a combination of constrained Hartree-Fock and exact diagonalization methods. 
We characterize the stability of the FQAH phase by the size of the FQAH gap and find that RnG generally has a stable FQAH phase, although the required displacement field varies significantly among different $n$ values. 
Our work establishes the theoretical universality of both IQAH and FQAH in RnG. 
\end{abstract}

\maketitle

\section{Introduction}
Recently, rhombohedral $n$-layer graphene (RnG) has stood out as a versatile and unique platform to realize various exotic phases of matter without the need for \moire\ superlattice, such as correlated insulating states~\cite{Chen2019,Zhou2021,Han2023,Winterer2024,Arp2024}, unconventional superconductivity~\cite{Chen2019a,Zhou2021a,Han2023a,Liu2023,Li2024}, and the integer quantum anomalous Hall (IQAH)~\cite{Chen2020,Han2024} effect. 
The past year has further witnessed the groundbreaking experimental observations of the fractional quantum anomalous Hall (FQAH) effect in R5G and R6G~\cite{Lu2024,XieJian2024}, which have since spurred intense theoretical interests in this field~\cite{Dong2023,Dong2023a,Zhou2023,Guo2023,Kwan2023,Zeng2024,Soejima2024,Dong2024,Tan2024,Huang2024,Yu2024,XieMing2024,Devakul2024}.
Although both IQAH and FQAH were predicted in flatband lattice systems without the application of an external magnetic field more than 10 years ago, no specific prediction was ever made for RnG systems, and therefore, these discoveries are unanticipated~\cite{Tang2011, Sun2011, Neupert2011, Regnault2011, Sheng2011}.

The current experimental realization of the FQAH effect (or FQAHE for short) in R5G and R6G is achieved under the following empirical conditions~\cite{Lu2024,XieJian2024}. 
First, the RnG is encapsulated between two layers of hexagonal boron nitride (hBN).  
Second, the RnG is aligned with the hBN on one side but misaligned on the other side, resulting in a \moire\ potential on the aligned layer of graphene. 
Finally, a strong perpendicular displacement field is applied to the sample, such that the electrons in the lowest conduction band are driven away from the strong \moire\ potential between the RnG and the hBN substrate.
Under such conditions, the IQAH effect (or IQAHE for short) and FQAHE were observed in R5G and R6G. 

Given the similarities between the noninteracting band structures of RnG for different numbers of layers $n$, it is natural to ask whether the FQAHE  can be observed in other RnG systems under similar conditions. 
Moreover, are there any trends in the appearance of the FQAHE in RnG with different $n$? 
In this work, we aim to address the above questions by presenting a theoretical comparison of the IQAHE and FQAHE in RnG in the presence of a strong displacement field within a unified and universal theoretical framework~\cite{Huang2024}. 
We will examine the stability of the IQAH and FQAH phases in RnG ($n=\numrange{3}{6}$) and thus establish the universality of IQAHE and FQAHE in RnG materials similar to the universality of continuum integer and fractional quantum Hall effects in two-dimensional electron gas systems in the presence of strong magnetic fields.

Our study is based on the comprehensive theory we developed in an earlier work~\cite{Huang2024} for the IQAHE and FQAHE in pentalayer graphene (R5G), motivated by Ref.~\cite{Lu2024}. 
In particular, our theory starts with the well-accepted noninteracting continuum model of RnG and then applies the Hartree-Fock (HF) approach to derive a quasiparticle band structure at the integer filling $\nu=1$, which is characterized by a flat lowest conduction band with a nonzero Chern number and is separated from other bands by a gap. 
The important features of our theory include the following. 
First, our theory includes a proper reference field in the HF calculation to ensure a convergent result within the momentum cutoff of the continuum model. 
Second, our theory incorporates all the valence bands within the momentum cutoff in the HF calculation to allow for an accurate calculation of the ground state energy. 
Third, we treat the HF calculation and the exact diagonalization (ED) as a unified framework to study the FQAHE in RnG. 
In particular, when searching for the ground states of FQAHE, we carry out the HF calculation directly at the desired fractional filling and then project the Hamiltonian to the resulting basis to perform the ED calculation. 

\begin{figure*}[!]
	\includegraphics[width=\textwidth]{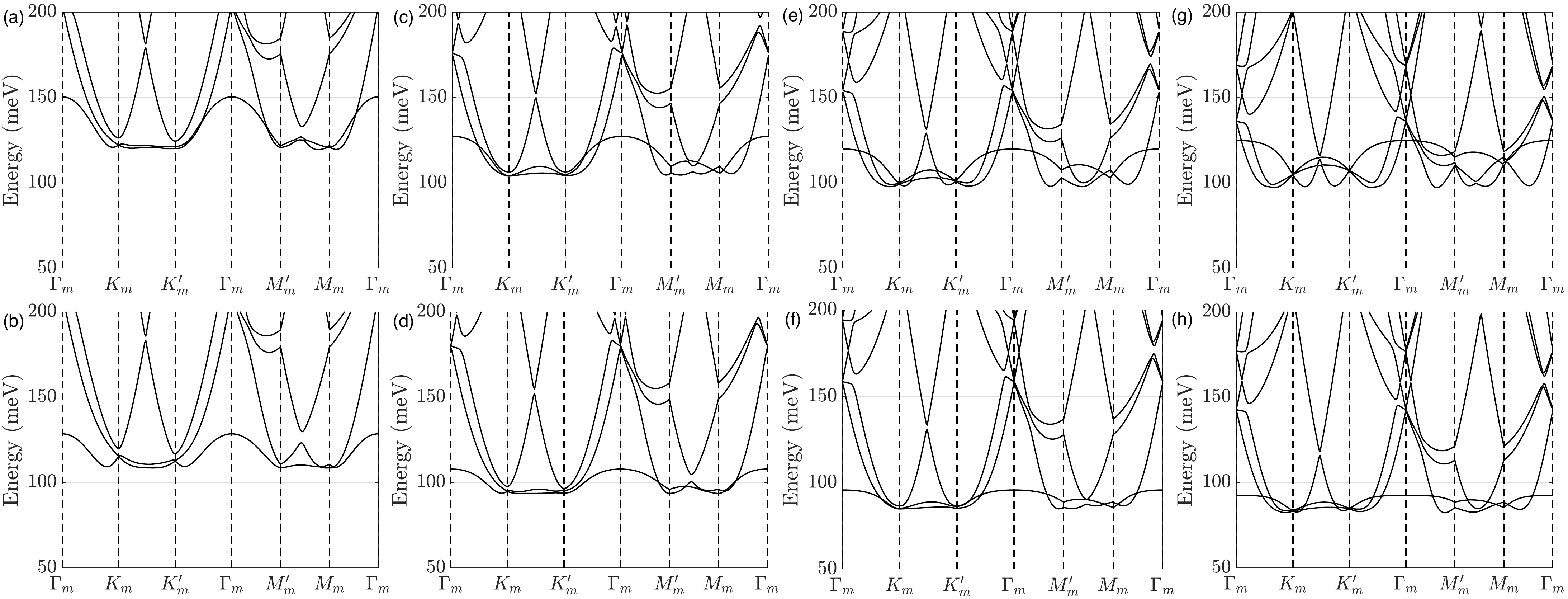}
	\center
	\caption{\label{Fig:Band}
	Noninteracting band structure of (a) R3G with $u_d=\SI{150}{meV}$, (b) R3G with $u_d=\SI{128}{meV}$, (c) R4G with $u_d=\SI{85}{meV}$, (d) R4G with $u_d=\SI{72}{meV}$, (e) R5G with $u_d=\SI{60}{meV}$, (f) R5G with $u_d=\SI{48}{meV}$, (g) R6G with $u_d=\SI{50}{meV}$, and (h) R6G with $u_d=\SI{37}{meV}$.
	Here, the twist angle is $\theta=0.77^\circ$.}
\end{figure*}

In the rest of the paper, we present our theoretical results on the IQAHE and FQAHE in RnG ($n=\numrange{3}{6}$) with a strong applied displacement field. 
We first present the noninteracting band structures and then apply the HF approximation to obtain the quasiparticle bands.
We then consider a fractional filling of the quasiparticle band and obtain the FQAH states using the ED method.

\section{Interacting model of RnG}
In the experiment, the RnG is aligned with one layer of hBN on one side but misaligned on the other side, resulting in a \moire\ potential on the aligned graphene layer. 
The noninteracting continuum model of spin $\sigma$ in such a system can be written as~\cite{Zhang2010,Jung2013,Moon2014},
\begin{align}
	H_{\text{s},\sigma}&=H_{\text{RnG}}(-i\nabla)+u_dV_d+V_{\text{\moire}}(r),
\end{align}
where the first term is the Hamiltonian of the pristine RnG, the second term is the perpendicular displacement field with $u_d$ denoting the voltage between the neighboring layers of the RnG, and the last term is the \moire\ potential acting only on the aligned layer. 
Their specific forms will be discussed below. 
Originating from the lattice mismatch and twist angle between RnG and hBN, the periodicity of the \moire\ potential determines the \moire\ Brillouin zone (MBZ) in the momentum space.


We now explain the specific forms of $H_{\text{RnG}}$, which captures the displacement field $V_d$, and the \moire\ potential $V_{\text{\moire}}$ in the RnG. 
In the momentum space, the Hamiltonian of the pristine RnG is modeled as
\begin{align}
H_{\text{RnG}}(\vb k)
 =\left[\begin{array}{ccccc}
 h^{(0)} & h^{(1)} & h^{(2)} &  & \\
 h^{(1)\dag} & h^{(0)} & \ddots & \ddots &  \\
 h^{(2)\dag} & \ddots & \ddots & h^{(1)} & h^{(2)} \\
  & \ddots & h^{(1)\dag} & h^{(0)} & h^{(1)} \\
  &  & h^{(2)\dag} & h^{(1)\dag} & h^{(0)}\end{array}\right],
\end{align}
where
\begin{align}
&h^{(0)}=-t_0\left[\begin{array}{cc}0 & f_{\vb k} \\ f_{\vb k}^* & 0\end{array}\right],\quad
h^{(1)}=\left[\begin{array}{cc}t_4f_{\vb k} & t_3 f_{\vb k}^* \\t_1 & t_4f_{\vb k} \end{array}\right],\nonumber\\
&h^{(2)}=\left[\begin{array}{cc}0 & \frac{t_2}{2} \\0 & 0\end{array}\right],\quad f_{\vb k}=\sum_{j}\exp(i\vb k\cdot \delta_j).
\end{align}
Following Ref.~\cite{Wang2024a}, we take 
\begin{align*}
	(t_0,t_1,t_2,t_3,t_4)=(3100,380,-21,290,141)\, \text{meV}. 
\end{align*}
Further, the displacement field is given by
\begin{align}
[V_{d}]_{ll'}=\delta_{ll'}[l-(n_l-1)/2],
\end{align}
where $l,l'$ are the indices of layer, and $n_l$ is the number of layers.
Here, $l=0$ represents the top layer, and $l=n_l-1$ represents the bottom layer.

\begin{figure*}[!]
	\includegraphics[width=\textwidth]{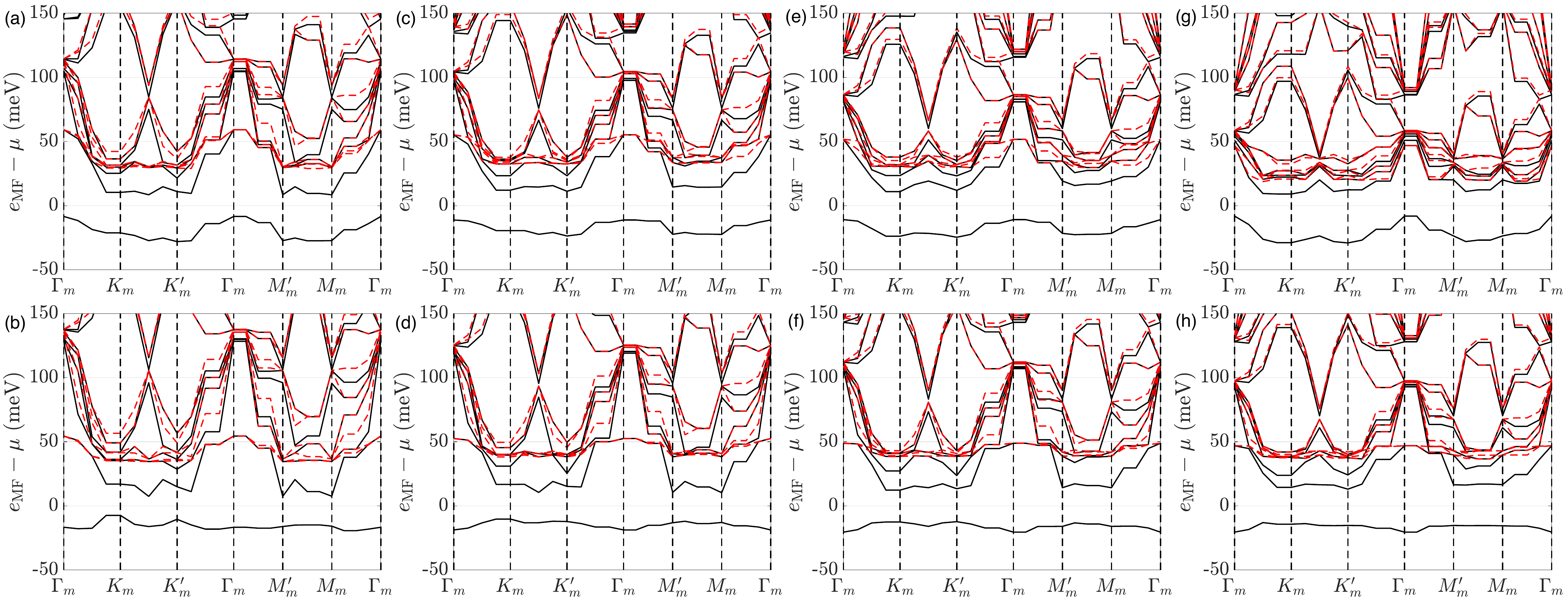}
	\center
	\caption{\label{Fig:HFBand}
	Noninteracting band structure of (a) R3G with $u_d=\SI{150}{meV}$, (b) R3G with $u_d=\SI{128}{meV}$, (c) R4G with $u_d=\SI{85}{meV}$, (d) R4G with $u_d=\SI{72}{meV}$, (e) R5G with $u_d=\SI{60}{meV}$, (f) R5G with $u_d=\SI{48}{meV}$, (g) R6G with $u_d=\SI{50}{meV}$, and (h) R6G with $u_d=\SI{37}{meV}$. The black lines denote the spin-up sector, and the red dashed lines denote the spin-down sector.  Here, the twist angle is $\theta=0.77^\circ$, and the calculation is performed on a $4\times6$ mesh in the MBZ with $S=\text{diag}[1,1]$.
	}
\end{figure*}

To model the \moire\ potential, we consider the aligned layer of RnG with hBN. 
At twist angle $\theta$, the reciprocal basis vectors of the aligned hBN are given by
\begin{align}
	\vb G_{i}'=\frac{a_\text{G}}{a_\text{hBN}}R_{\theta}\vb G_i, \quad (i=1,2), 
\end{align}
with a lattice mismatch $a_\text{hBN}/a_\text{G}\approx 1.018$. 
Here $\vb G_i$ are the reciprocal lattice vectors of graphene. 
Hence, the \moire\ reciprocal vectors are given by the difference between the reciprocal vectors of the RnG and hBN, i.e., $\vb g_i=\vb G_i-\vb G_i'$ for $i=1,2$ and $\vb g_3=-\vb g_1-\vb g_2$.
In this work, we follow Ref.~\cite{Moon2014} and take the following form for the \moire\ potential on the aligned layer in the $\vb K$ valley,
\begin{align}
	V_{\text{\moire}}(\vb r)=V_0\mathbb{I}+V_1\sum_{j=1}^3\qty[ e^{i\psi} U_j e^{i\vb r\cdot\vb g_j}+\text{H.c.}],
\end{align}
where
\begin{align*}
	U_1=\left[\begin{array}{cc} 1 & 1 \\ \omega & \omega\end{array}\right],\quad 
	U_2=\left[\begin{array}{cc} 1 & 1/\omega \\ 1/\omega & \omega\end{array}\right],\quad
	U_3=\left[\begin{array}{cc} 1 & \omega \\ 1  & \omega\end{array}\right],
\end{align*}
with $\omega=e^{2\pi/3 i}$, $\mathbb{I}$ is identity, and $(V_0,V_1,\psi)=(28.9\,\text{meV},21\,\text{meV},0.29)$. 
The $\vb K'$ valley is defined by the time-reversal symmetry.

For the many-body interaction, we consider the gate-screened Coulomb interaction, given by
\begin{align}
	V_{\text{Coulomb}}=\frac{1}{2A}\sum_{\vb q}\dfrac{\tanh(dq)}{2\epsilon\epsilon_0 q}:\rho_{-\vb q}\,\rho_{\vb q}:,
\end{align}
where $A$ is the area of the system, $:\mathcal O:$ denotes the normal order of an operator, $d$ is the distance between the sample and the gate (taken to be $\SI{100}{nm}$), and $\epsilon$ is the background dielectric constant (taken to be $5$ in this work). 
Here, we define the density operator $\rho_{\vb q}=\sum_{\alpha,\sigma}\sum_{\vb k}c_{\vb k+\vb q,\alpha,\sigma}^\dag c_{\vb k,\alpha,\sigma}$, and $c_{\vb k,\alpha,\sigma}$ denotes the annihilation operator of a plane wave state, where $\alpha$ represents the collective index of layer, sublattice, and valley. 
The total Hamiltonian is then given by
\begin{align}
	H=\sum_{\sigma}H_{\text{s},\sigma}+V_{\text{Coulomb}}-V_{\text{HF}}(P_\text{ref}),
\end{align}
where the third term is induced to avoid double counting the interaction within the momentum cutoff of the continuum model. 
$V_{\text{HF}}(P)$ denotes the Coulomb interaction in the HF approximation and is a linear functional of the one-body density matrix $P$. 
$P_{\text{ref}}$ here is the so-called ``reference field'' or ``subtraction scheme'', which does not have an accepted form in the literature. 
Because of the convergence issue discussed in Ref.~\cite{Huang2024}, we take $P_{\text{ref}}$ to be the noninteracting ground state at the charge neutrality point. 
This choice of reference field is known in the literature as the charge neutrality scheme, and other possible choices are discussed in Ref.~\cite{Huang2024}.

\begin{figure*}[t]
	\includegraphics[width=\textwidth]{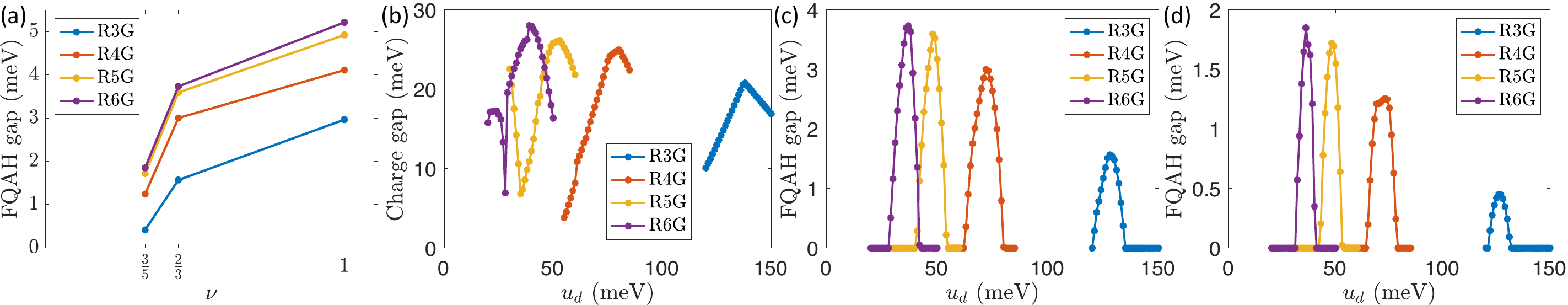}
	\center
	\caption{\label{Fig:FCI_135}
	(a) FQAH gap at various filling factors. We take $u_d=\SI{128}{meV},\SI{72}{meV},\SI{48}{meV},\SI{37}{meV}$ for R3G, R4G, R5G, and R6G, respectively. For $\nu=1$, we show the charge gaps in the HF calculation at $\nu=1$, and we write $\times5$ for those data points, meaning that the actual gaps are five times as large as the data points in the plot. (b) Charge gap at $\nu=1$, (c) FQAH gap at $\nu=2/3$, and (d) FQAH gap at $\nu=3/5$ as a function of the displacement field. In all panels, the twist angle is $\theta=0.77^\circ$, and the calculation is performed on a $4\times6$ mesh in the MBZ for $\nu=1,2/3$ and on a $5\times5$ mesh in the MBZ for $\nu=3/5$ with $S=\text{diag}[1,1]$ for both meshes.
	}
\end{figure*}

\section{The IQAHE in RnG}
The experimental observation of the IQAHE and FQAHE is theoretically baffling because the noninteracting band structure of the RnG with a strong displacement field is highly entangled and with no discernible gaps in the spectrum, as shown in Fig.~\ref{Fig:Band}. 
Moreover, we observe that the stronger the displacement field, the more entangled the conduction bands. 
Hence, a naive noninteracting band structure cannot explain the IQAHE, let alone the FQAHE. 
To resolve this problem, an HF approach is proposed to understand the appearance of the IQAH~\cite{Dong2023,Dong2023a,Zhou2023,Kwan2023,Yu2024}, in which an HF charge gap is formed associated with a spontaneous symmetry breaking leading to a Chern number, separating the lowest HF conduction band from the rest. 
The numerical calculation necessitates discrete meshes in the MBZ, and we consider the following $N_1\times N_2$ momentum mesh,
$\vb k=  n_1/N_1 \tilde{\vb g}_1 +  n_2/N_2 \tilde{\vb g}_2$, where $n_i$ takes value from $0$ to $N_i-1$, $\tilde{\vb g}_i=\sum_{j=1}^2S_{ij}\vb g_j$, and $S_{ij}$ is an integer-valued matrix with $\det S=1$. 
Here, $S$ amounts to changing the basis vectors of the reciprocal lattice.

In Fig.~\ref{Fig:HFBand}, we calculate the HF band structure at filling factor $\nu=1$ in RnG ($n=\numrange{3}{6}$) for two values of $u_d$. In this work, we always include all the bands within the momentum cutoff in the HF calculation instead of projecting the system to a few valence and conduction bands. 
For the parameters studied in Fig.~\ref{Fig:HFBand}, we find that the spin-valley polarized solutions are always energetically favored in the HF calculation and that a correlated gap is always formed between the lowest conduction band and the other conduction bands.
Furthermore, there is a sweet spot for each of these structures, respectively, where the lowest conduction band becomes considerably flatter, as shown in the bottom row of Fig.~\ref{Fig:HFBand}. 
Moreover, the HF band is topologically nontrivial with Chern number $C=1$. 
However, the bandwidth of the lowest quasiparticle conduction band increases significantly as the displacement field deviates from this sweet spot, as shown in the top row of Fig.~\ref{Fig:HFBand}. 
This clearly establishes the necessity for the applied electric field in obtaining IQAHE and FQAHE in RnG.

\section{The FQAHE in RnG}
While the HF calculation can predict the IQAHE at integer fillings, it fails to capture the strongly correlated ground states at fractional fillings,  although it has been hypothesized~\cite{Sun2011} that a partially filled flat Chern band could result in the FQAH states. 
However, this hypothesis must be verified explicitly for each specific case.
One of the most reliable methods to investigate the FQAH states is the exact diagonalization (ED), which is limited to one or two bands. 
However, there are multiple highly entangled bands in the RnG, far beyond the capability of the ED.
Notwithstanding, if one assumes that the electron correlation only appears in the lowest quasiparticle conduction band, the Hamiltonian can be projected to the following subspace~\cite{Huang2024}
\begin{align}\label{Eq:Ansatz}
	\mathcal P=\text{span}\qty\Big{\prod_{i=1}^{N} f_{c,\vb k_i}^\dag\prod_{\alpha,\vb k}{f_{v,\alpha,\vb k}^\dag}\ket{0}: \vb k_i\in\text{MBZ}},
\end{align}
where $f_{c,\vb k}$ is the partially filled quasiparticle conduction band, and $f_{v,\alpha,\vb k}$ is the completely filled quasiparticle valence bands. 
In principle, the quasiparticle bands should be determined variationally by minimizing the ground state energy. 
However, this is numerically impossible. 
In practice, one can only compare the energies of different candidate sets of quasiparticle bands, and the quasiparticle bands derived from a constrained HF calculation at the corresponding fractional filling are found to produce the most energetically favorable FQAH ground states~\cite{Huang2024}. 
Essentially, the constrained HF calculation requires the one-body density matrix to have a uniform density in the MBZ, which is a hallmark of the FQAH states. 
This method is equivalent to treating the interband interaction on the HF level but treating the intraband interaction within the quasiparticle conduction band on the ED level to generate an FQAH state.

\begin{figure}[t]
	\includegraphics[width=\columnwidth]{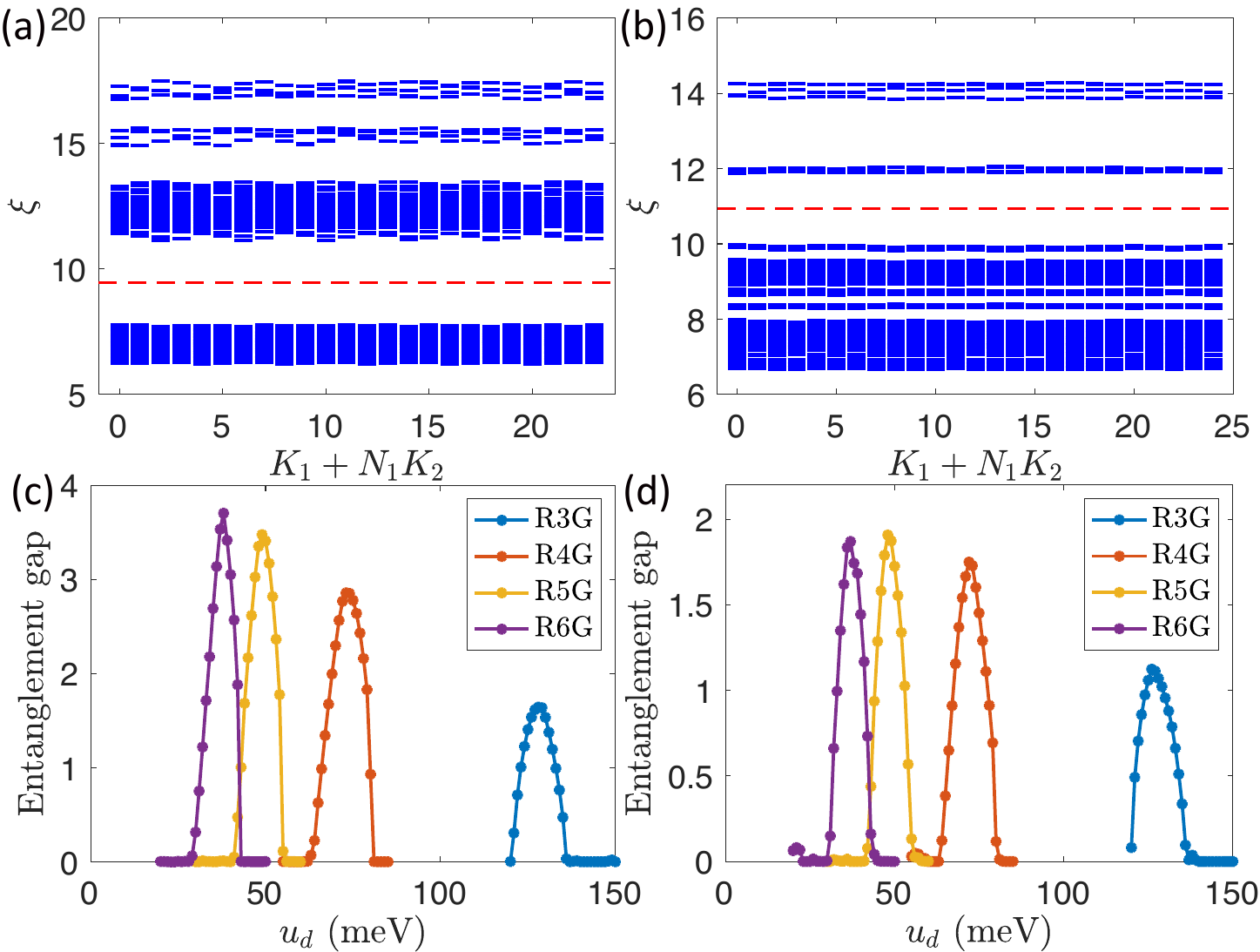}
	\center
	\caption{\label{Fig:EE}
	Upper panel: Momentum-resolved PES of the particle-hole conjugate of the ground state in R5G at (a) $\nu=2/3$ and (b) $\nu=3/5$. Here, we take with $\theta=0.77^\circ$ and $u_d=\SI{48}{meV}$. There are 1088 and 2150 states below the red dashed lines in (a) and (b), respectively, as per the state counting of the FQAH states~\cite{Regnault2011}.
	Lower panel: FQAH entanglement gap at (c) $\nu=2/3$ and (d) $\nu=3/5$ as a function of the displacement field. 
	Note that the gap in (c) and (d) refers to the one marked by the red dashed line in (a) and (b), respectively. Therefore, a nonzero gap in (c) and (d) indicates the existence of the FQAH phase. 
	We use the same parameters as those in Fig.~\ref{Fig:FCI_135}, and trace out $N_a=5$ and $N_a=7$ particles for $\nu=2/3$ and $\nu=3/5$, respectively. 
	}
\end{figure}

\begin{figure*}[t]
	\includegraphics[width=\textwidth]{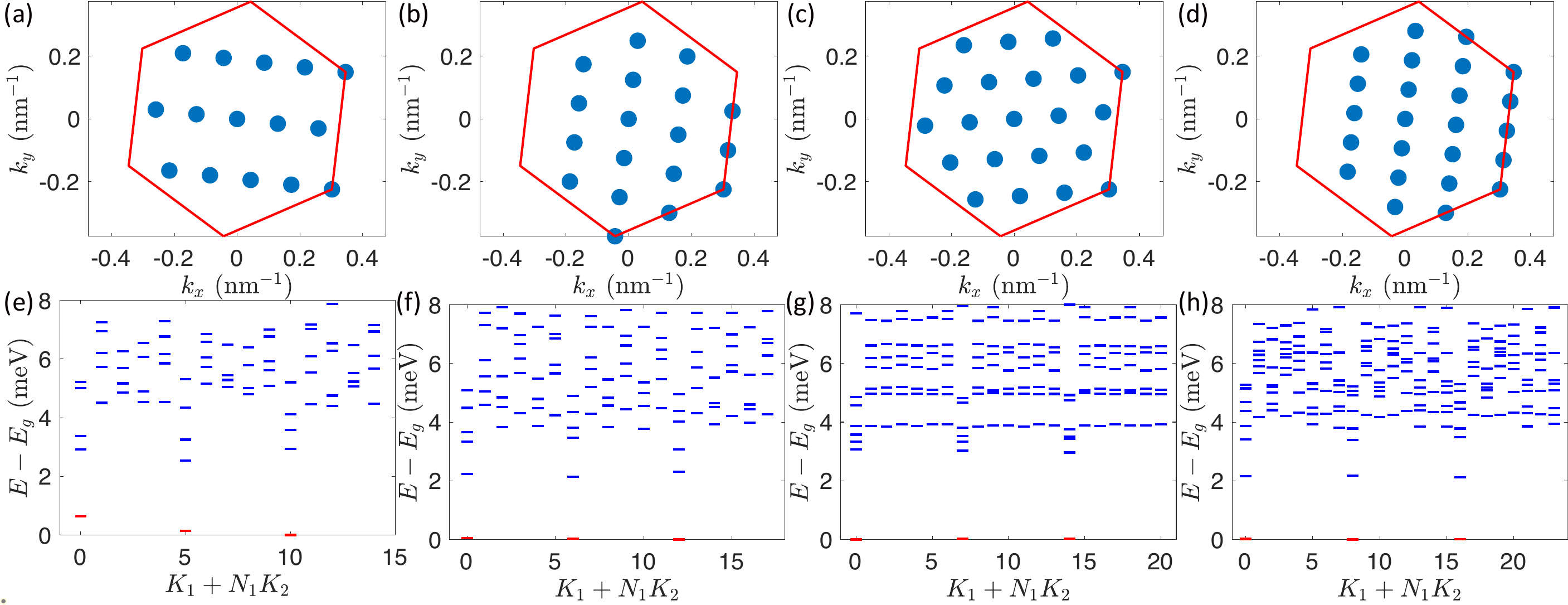}
	\center
	\caption{\label{Fig:mesh}
	Upper panels: Momentum meshes for (a) $N_1\times N_2=1\times15$ with $S=\left[\begin{array}{cc}1 & -5 \\ 1 & -4\end{array}\right]$, (b) $N_1\times N_2=2\times9$ with $S=\left[\begin{array}{cc}1 & -2 \\ 1 & -1\end{array}\right]$, (c) $N_1\times N_2=1\times21$ with $S=\left[\begin{array}{cc}1 & -5 \\ 1 & -4\end{array}\right]$, and (d) $N_1\times N_2=4\times6$ with $S=\left[\begin{array}{cc}1 & 1 \\ 1 & 2\end{array}\right]$. The red lines delineate the boundary of the first MBZ. Lower panels: energy spectra in each momentum sector for R5G at $\nu=2/3$ using the corresponding momentum meshes in the upper panels. The red lines label the three-fold degenerate ground states, and $E_g$ is the ground state energy. Here, the twist angle is $\theta=0.77^\circ$ and $u_d=\SI{48}{meV}$.
	}
\end{figure*}

We now study the FQAH state in RnG ($n=\numrange{3}{6}$) using a combination of the constrained HF and ED calculations described above. 
The results are presented in Fig.~\ref{Fig:FCI_135}. 
Here, we characterize the stability of the FQAH phase by the FQAH gap. 
As the FQAH state follows the generalized Pauli exclusion principle, the degenerate ground states only appear in the dictated momentum sectors~\cite{Regnault2011}. 
Therefore, the FQAH gap is defined as the energy gap between the highest ground state energy in the dictated momentum sectors and the rest of the energies. 
In particular, the FQAH gap is set to zero if such a gap does not exist. 
For the displacement field used in the upper panels of Fig.~\ref{Fig:Band} and Fig.~\ref{Fig:HFBand}, we find that the FQAH gap vanishes in the corresponding system because of the large dispersion in the quasiparticle conduction band. 
In contrast, for the displacement used in the lower panels of Fig.~\ref{Fig:Band} and Fig.~\ref{Fig:HFBand}, we find a finite FQAH gap at both $\nu=2/3$ and $\nu=3/5$, as shown in Fig.~\ref{Fig:FCI_135}(a). 
We also note that the FQAH gap at $\nu=2/3$ is larger than that at $\nu=3/5$ but much smaller than the charge gap at $\nu=1$.

Finally, we study the energy gaps as a function of the displacement field at different fillings. 
At filling factor $\nu=1$, the HF calculation always opens up a gap for the parameter range we studied, as shown in Fig.~\ref{Fig:FCI_135}(b). 
However, we indeed observe a competition between HF solutions with different Chern numbers in R5G and R6G, and their transition results in cusps at $u_d=\SI{35}{meV}$ for R5G and $u_d=\SI{28}{meV}$ for R6G. 
At $2/3$ filling, all four systems have a finite region of FQAH states, as shown in Fig.~\ref{Fig:FCI_135}(c). 
We also notice that the necessary displacement field in the R4G to generate FQAH states is stronger than in the R5G. 
This agrees with the experimental observations, although the experiments of R4G and R5G are performed at two slightly different twist angles~\cite{LongJu}. 
Meanwhile, the displacement field required to generate FQAH states in R3G is significantly larger (this fact may have prevented the experimental manifestation of the FQAHE in R3G.)
We also examine the FQAHE at $3/5$ filling. 
The nonvanishing FQAH gap shown in Fig.~\ref{Fig:FCI_135}(d) suggests that the FQAH states still exist. 
However, compared to the $2/3$ filling case, both the range of the FQAH phase and the magnitude of the nonvanishing FQAH gap are smaller, confirming the generic well-known trend that FQHE is less stable for higher-order fractional states.

Another defining feature of the FQAH state is the entanglement gap in the particle entanglement spectrum (PES). The PES is defined as the spectrum of $-\ln[\trace_{N_a}(\rho)]$, where $\rho$ is the ground state density matrix, and $\trace_{N_a}$ denotes tracing out $N_a$ particles. For the FQAH state, we take $\rho=\sum_{j=1}^3\dyad{\psi_j}/3$, where $\ket{\psi_j}$ are the three-fold degenerate ground states. 
As the PES probes the quasiparticle excitation of the ground state, the FQAH state has an entanglement gap, below which the number of states follows the generalized Pauli principle~\cite{Regnault2011}.
In Fig.~\ref{Fig:EE}, we calculate the entanglement gap of the particle-hole conjugate of the ground state in RnG ($n=\numrange{3}{6}$) at $\nu=2/3$ and $\nu=3/5$. 
The entanglement gap agrees with the energy gap and further verifies the existence of the FQAH phase
in RnG.

\section{Different system sizes and momentum meshes}

\begin{figure}[t]
	\includegraphics[width=\columnwidth]{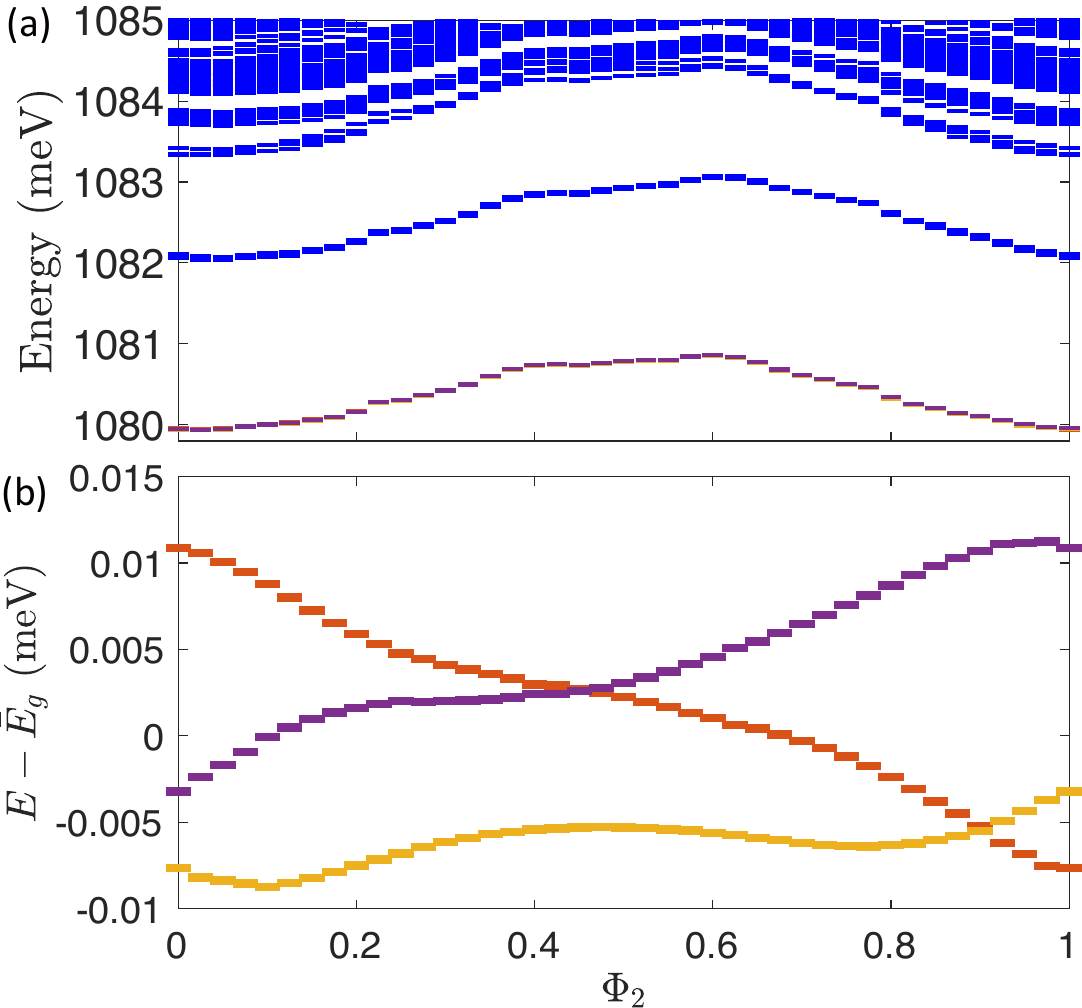}
	\center
	\caption{\label{Fig:flux}
	(a) Energy spectrum of R5G at $\nu=2/3$ as a function of inserted flux $\Phi_2$. (b) Permutation of the three-fold degenerate ground states after inserting one unit of flux. The red, purple, and yellow lines respectively denote the three-fold degenerate state in the three momentum sectors, and $\bar E_g$ is the mean energy of the three ground states. Here, we use the same parameters as those in Fig.~\ref{Fig:mesh}(d) and~\ref{Fig:mesh}(h).
	}
\end{figure}

In the previous section, we demonstrate the FQAH states on two momentum meshes, respectively, for two fractional fillings. 
However, the size and shape of the momentum mesh may affect the result in finite-size systems. 
Particularly, it was argued that the high-symmetry points may play an indispensable role in the formation of the FQAH state~\cite{Soejima2024}. In this section, we focus on the R5G at $\nu=2/3$ and study the results for different momentum meshes with $N_1\times N_2=15,18,21,24$, as shown in Fig.~\ref{Fig:mesh}(a-d). Specifically, Fig.~\ref{Fig:mesh}(a) and contains $\Gamma_m$, and two $K_m$; Fig.~\ref{Fig:mesh}(b) contains $\Gamma_m$, two $K_m$, and one $M_m$; Fig.~\ref{Fig:mesh}(c) contains $\Gamma_m$, two $K_m$, and respects the $C_3$ symmetry; Fig.~\ref{Fig:mesh}(d) contains all the high-symmetry points. Despite different system sizes and meshes, we observe in Fig.~\ref{Fig:mesh}(e-h) that the FQAH energy gap remains finite.
Another way to deform the mesh is to shift it in the momentum space, i.e., $\vb k \to \vb k+\Phi_2  \tilde{\vb g}_2/N_2$. This is equivalent to inserting fluxes into the system~\cite{Regnault2011}, and a stable FQAH state requires that the FQAH energy gap remains open during the flux insertion. In Fig.~\ref{Fig:flux}(a), we calculate the energy spectrum of R5G at $\nu=2/3$ versus the flux insertion, using the momentum mesh given in Fig.~\ref{Fig:mesh}(d). The FQAH gap is almost independent of the flux, and the energy splitting among the three-fold ground states is always very small. 
However, we find that there is an overall energy shift dependent on the flux, because we include the energy of the valence band in the calculation, which depends on the momentum meshes in the HF calculation.
Nonetheless, the energy shift is about $\SI{1}{meV}$ and fairly small compared to the total energy of the state. {We further show in Fig.~\ref{Fig:flux}(b) that the three-fold degenerate states flow into one another after inserting one unit of flux. However, we note that the spectral flow in a $N_1\times N_2=4\times6$ system should be construed as a corollary of the generalized Pauli principle rather than an additional signature of FCI. If the system size is not divisible by 9, then the generalized Pauli principle dictates that three degenerate ground states are in different momentum sectors. 
Particularly, if $N_2$ is divisible by $3$ but $N_1$ is not, the $K_2$ component of the total momentum of the ground states without flux is given by $K_2/N_2=0$, $1/3$, and $2/3$. Inserting a flux along $\tilde{\vb g}_2$ increases the momentum of each particle by $\Phi_2/N_2$, and thus, the total momentum becomes $K_2/N_2=\nu \Phi_2 N_1,\nu \Phi_2 N_1+1/3,\nu \Phi_2 N_1+2/3$, implying that the three momentum sectors are adiabatically connected to one another. 
As $N_1$ is not divisible by $3$, the three momentum sectors are guaranteed to experience a cyclic permutation after inserting one unit of flux.}

\section{Discussion and conclusion}
In this work, we have studied the IQAHE and FQAHE in RnG with a strong displacement field. 
Our study aims to apply the comprehensive theory we developed previously for the IQAHE and FQAHE in R5G to other RnG systems and examine the stability of the FQAH phase in these systems. 
Our results show that under similar conditions, a quasiparticle conduction band with $C=1$ can be formed in RnG at the integer filling $\nu=1$, leading to the IQAHE. 
Moreover, upon hole doping the lowest quasiparticle conduction band, the FQAHE can be observed at certain fractional fillings in RnG in general. 
Despite this general trend, several interesting features are observed in the FQAHE in RnG. 
First, the FQAH gap at $\nu=2/3$ and $\nu=3/5$ generally decreases with the number of layers $n$. 
Second, the center of the FQAH phase in the displacement field axis is shifted to a larger value as $n$ decreases. 
This suggests that a very large displacement field may be required to observe the FQAHE in R3G. 
However, a quantitative comparison between the FQAH gaps in different RnG systems is currently challenging because of the different parameters used in the experiments, such as the twist angle between the RnG and hBN for different $n$. 
In addition, current experiments appear to have large background disorder effects~\cite{XieMing2024}, making any comparison between theory and experiment impossible since strong disorder would generically suppress both FQAHE and IQAHE~\cite{Yang2012}. 

\section{Acknowledgements}
K.H. and X.L. are supported by the Research Grants Council of Hong Kong (Grants No. CityU 11300421, CityU 11304823, and C7012-21G) and City University of Hong Kong (Projects No. 9610428 and 7005938). 
K.H. is also supported by the Hong Kong PhD Fellowship Scheme. 
S.D.S. is supported by the Laboratory for Physical Sciences through the Condensed Matter Theory Center (CMTC) at the University of Maryland. 
This work was performed in part at the Aspen Center for Physics, which is supported by National Science Foundation grant PHY-2210452.

\bibliography{RnG_bib}

\end{document}